\documentclass[12pt,preprint]{aastex}

\slugcomment{Submitted to The Astrophysical Journal}

\lefthead{}
\righthead{}

\begin{document}

\title{THREE-DIMENSIONAL MHD SIMULATIONS OF RADIATIVELY INEFFICIENT
ACCRETION FLOWS}

\author{Igor V. Igumenshchev}

\affil{Laboratory for Laser Energetics, University of Rochester, 250
East River Road, Rochester, NY 14623; iigu@lle.rochester.edu}

\author{Ramesh Narayan}

\affil{Department of Astronomy, Harvard-Smithsonian Center for
Astrophysics, 60 Garden Street, Cambridge, MA 02138;
narayan@cfa.harvard.edu}

\author{Marek A. Abramowicz}

\affil{Department of Astronomy and Astrophysics, G\"oteborg
University and Chalmers University of Technology, S-41296, G\"oteborg,
Sweden; marek@fy.chalmers.se}

\begin{abstract}

We present three-dimensional MHD simulations of rotating radiatively
inefficient accretion flows onto black holes.  We continuously inject
magnetized matter into the computational domain near the outer
boundary and run the calculations long enough for the resulting
accretion flow to reach a quasi-steady state.  We have studied two
limiting cases for the geometry of the injected magnetic field: pure
toroidal field and pure poloidal field.  In the case of toroidal field
injection, the accreting matter forms a nearly axisymmetric,
geometrically-thick, turbulent accretion disk.  The disk resembles in
many respects the convection-dominated accretion flows found in
previous numerical and analytical investigations of viscous
hydrodynamic flows.  Models with poloidal field injection evolve
through two distinct phases.  In an initial transient phase, the flow
forms a relatively flattened, quasi-Keplerian disk with a hot corona
and a bipolar outflow.  However, when the flow later achieves steady
state, it changes in character completely.  The magnetized accreting
gas becomes two-phase, with most of the volume being dominated by a
strong dipolar magnetic field from which a thermal low-density wind
flows out.  Accretion occurs mainly via narrow slowly-rotating radial
streams which `diffuse' through the magnetic field with the help of
magnetic reconnection events.

\end{abstract}

\keywords{accretion, accretion disks --- black hole physics ---
convection --- MHD --- turbulence}

\section{INTRODUCTION}

This study is a continuation of our efforts during the last several
years to understand the nature of underluminous accreting black holes.
A class of radiatively inefficient solutions (Ichimaru 1977; Rees et
al. 1982; Narayan \& Yi 1994, 1995a,b; Abramowicz et al. 1995;
Narayan, Mahadevan \& Quataert 1998) has been influential in this
field.  A key feature of these solutions is that radiative energy
losses are small so that most of the energy is advected with the gas.
Narayan \& Yi (1994) came up with the name `advection-dominated
accretion' to describe such flows, and Lasota (1996) suggested the
acronym ADAF (advection-dominated accretion flow).  Usually by ADAFs
one understands optically thin accretion flows with very low mass
accretion rates, $\dot{M}\ll \dot M_{Edd}$, but ADAFs are also
possible as optically thick flows with high accretion rates, $\dot{M}
> \dot M_{Edd}$, where $\dot M_{Edd}$ is the Eddington accretion
rate. In the later case, one uses the name `thick accretion disk' (see
e.g. Jaroszy{\'n}ski, Abramowicz \& Paczy{\'n}ski, 1980) or `slim
accretion disk' (Abramowicz et al., 1988) rather than ADAF.

Historically, ADAFs were initially studied within the framework of
viscous hydrodynamics.  In this approach, one assumes that the
accreting gas has an `anomalous viscosity' parameterized by the
standard dimensionless viscosity parameter $\alpha$.  Early analytic
work revealed that the gas in an ADAF would be (i) gravitationally
unbound because of a positive Bernoulli parameter, which might lead to
a strong outflow (Narayan \& Yi 1994, 1995a), and (ii) convectively
unstable (Gilham 1981; Begelman \& Meier 1982; Narayan \& Yi 1994,
1995a).  The important role of these processes was not, however,
realized until numerical two- and three-dimensional hydrodynamic
simulations were carried out by a number of authors (Igumenshchev,
Chen \& Abramowicz 1996; Igumenshchev \& Abramowicz 1999, 2000; Stone,
Pringle \& Begelman 1999; Igumenshchev, Abramowicz \& Narayan 2000;
McKinney \& Gammie 2002).  These studies showed that the behavior of a
viscous ADAF depends on the value of $\alpha$.  If $\alpha$ is large,
say $>0.3$, then there is a powerful bipolar outflow driven by the
positive Bernoulli parameter.  However, if $\alpha$ is small, say
$<0.1$, the flow is dominated by convective motions which strongly
modify the structure of the flow.

The convection-dominated low-$\alpha$ case was modeled analytically by
Narayan, Igumenshchev \& Abramowicz (2000) and Quataert \& Gruzinov
(2000), who coined the name `convection-dominated accretion flow'
(CDAF).  Two key features of a CDAF are: (i) the Reynolds stress
associated with the convective turbulence moves angular momentum
inward rather than outward, and (ii) the density profile is much
flatter than in the case of a pure ADAF --- $\rho \sim R^{-1/2}$
rather than $R^{-3/2}$.  The CDAF model provides a physical
explanation for the low luminosity of accreting black holes, namely,
the mass accretion rate is much reduced in the presence of convection.

Accretion models that are based on the principles of viscous
hydrodynamics are limited by the fact that they need to invoke a
mysterious `anomalous viscosity.'  Since the `viscosity' in these
models is very likely provided by magnetic fields via the
magneto-rotational instability (Balbus \& Hawley 1991), a deeper
understanding of accretion flows can clearly be obtained by analyzing
the problem directly within the framework of MHD.  Numerical MHD
simulations of radiatively inefficient flows have been carried out
recently by a number of authors (Hawley 2000, 2001; Stone \& Pringle
2001; Machida, Matsumoto \& Mineshige 2001; Hawley, Balbus \& Stone
2001; Hawley \& Balbus 2002; Casse \& Keppens 2002).  However, the
results obtained by the various authors have not been entirely
consistent.  For instance, some authors conclude that radiatively
inefficient MHD flows behave like CDAFs (Machida et al. 2001), whereas
others claim that the flows are very different (Hawley et al. 2001).
It is not clear if the discrepancies arise from differences in the
assumed geometry of the initial magnetic field or differences in the
numerical techniques. In addition, magnetic stresses can have
explicitly non-viscous behavior, such as causing angular momentum
transport and accretion without any associated energy dissipation (Li
2003).

We present in this paper 3D numerical simulations that we have carried
out of radiatively inefficient MHD accretion flows.  In \S2 we
describe our numerical method and the initial and boundary
conditions. In \S3 we present numerical results for two topologies of
the injected magnetic field: toroidal field (\S3.1) and poloidal field
(\S3.2).  In \S4 we discuss the implications of our results.  We argue
that the toroidal field case has many points of similarity with the
CDAF model (\S4.1), while the poloidal field case behaves very
differently (\S4.2).  In fact, the steady state configuration that we
obtain for the poloidal problem is unlike anything published
previously in the literature; we suggest possible reasons for this
(\S4.3) and discuss the applicability of the obtained solutions to
real accretion flows (\S4.4).  In \S5 we conclude with a summary.

\section{SIMULATION TECHNIQUE}

\subsection{Magnetohydrodynamic Equations and Numerical Method}

We consider the dynamics of accreting plasma within the framework of
the one-fluid MHD approximation. The relevant equations of resistive
MHD take the following form:
\begin{equation}
{d\rho\over dt} + \rho{\bf\nabla\cdot v} = 0,
\end{equation}
\begin{equation}
\rho{d{\bf v}\over dt} = -{\bf\nabla} (P_g+Q) - \rho{\bf\nabla}\Phi +
{1\over 4\pi}({\bf\nabla}\times{\bf B})\times {\bf B},
\end{equation}
\begin{equation}
\rho{d\epsilon\over dt} = -(P_g+Q){\bf\nabla\cdot v} + {1\over
4\pi}\eta{\bf J}^2,
\end{equation}
\begin{equation}
{\partial{\bf B}\over \partial t} = {\bf\nabla}\times({\bf v}\times{\bf B} -
\eta{\bf J}),
\end{equation}
where $\rho$ is the density, ${\bf v}$ is the velocity, $P_g$ is the
gas pressure, $\Phi$ is the gravitational potential, ${\bf B}$ is the
magnetic induction, $\epsilon$ is the specific internal energy, ${\bf
J}={\bf\nabla}\times{\bf B}$ is the current density, $\eta$ is the
resistivity, and $Q$ is the additional `pressure' corresponding to a
standard artificial viscosity.  

We adopt the ideal gas equation of state,
\begin{equation}
P_g=(\gamma-1)\rho\epsilon,
\end{equation}
with an adiabatic index $\gamma=5/3$, and we assume that there is no
radiative cooling.  We neglect self-gravity of the gas and take the
gravitational potential to be entirely due to a central black hole of
mass $M$.  We employ a pseudo-Newtonian approximation (Paczy\'nski \&
Wiita 1980) for the potential:
\begin{equation}
\Phi=-{GM\over R-R_g}, \qquad R_g={2\,GM\over c^2},
\end{equation}
where $R$ measures the radial distance from the black hole and $R_g$
is the gravitational radius of the black hole.

Our numerical method is the MHD generalization of the hydrodynamic PPM
algorithm (Colella \& Woodward 1984) and is identical to that used by
Igumenshchev \& Narayan (2002).  In this method the hydrodynamical
part is solved by means of a variant of the Godunov method and the
magnetic part is solved using a modified version of the method of
characteristics (Stone \& Norman 1992).  For reasons of numerical
stability, we solve an equation for the internal energy rather than
total energy.  This requires use of the term $Q$ in equations (2) and
(3), and the terms $\eta{\bf J}^2/4\pi$ and $\eta{\bf J}$ 
in equations (3) and (4), respectively.  To avoid excessive
loss of energy as a result of numerical reconnection of magnetic field
lines, the resistivity is assumed to be non-zero and to have the
following form (see Stone \& Pringle 1999),
\begin{equation}
\eta=\eta_0{|{\bf\nabla}\times{\bf B}|\over\sqrt{4\pi\rho}}\Delta^2,
\end{equation}
where $\eta_0$ is a dimensionless parameter and $\Delta$ is the grid
spacing. 

The code operates on a three-dimensional nested Cartesian grid, which
is designed so as to adequately resolve the large range of spatial
scales spanned by the accretion flow.  We designate the distance from
the black hole to the face centers of the outermost cube of the grid
as $R_{out}$.  Igumenshchev \& Narayan (2002) discuss the code in more
detail.

\subsection{Initial and Boundary Conditions}

The simulations are begun with non-magnetized `empty space'
surrounding the black hole; for technical reasons the empty space
corresponds in actuality to a non-zero but extremely low density of
matter.  Mass is steadily injected into the computational domain
inside a slender equatorial torus of thickness equal to about two
grid-cells centered on the black hole and with a radius nearly equal
to $R_{out}$.  
The injected matter has Keplerian rotation velocity and
internal energy equal to a fraction $0.2$ of the gravitational energy.
The matter also has a magnetic field associated with it.  We have
adopted two configurations for the injected field: purely toroidal
field and purely poloidal field (defined with respect to the
equatorial plane of the torus).

In the toroidal injection case, the field is generated assuming that
only the vertical component $A_z$ of the vector potential ${\bf A}$ is
non-zero in the injected matter; the other two components, $A_r$,
$A_\phi$, are set to zero. Here and below, we represent ${\bf A}$ in
cylindrical coordinates $(r,\phi,z)$. In each time step, $A_z$ is
increased within the torus as follows,
\begin{equation}
A_z^{\rm new}=A_z^{\rm old}+
\left({8\pi\Delta\rho c_s^2\over\beta_0}\right)^{1/2} \Delta,
\end{equation}
where $\Delta\rho$ is the increase of the density in the torus due to
the injection of matter, $c_s^2$ is the sound speed of the injected
matter, and $\Delta$ is the grid spacing.  The parameter $\beta_0$
measures the plasma $\beta$ of the injected material ($\beta\equiv
P_g/P_m$ where $P_m$ is the magnetic pressure).  Since the injection
of matter is axisymmetric, this model generates magnetic field with
closed axisymmetric toroidal magnetic lines. In reality, however, the
finite resolution of the Cartesian grid introduces small
non-axisymmetric perturbations to the field structure.  These act as
seed perturbations in the simulations.  Since ${\bf
B}=\nabla\times{\bf A}$, simply increasing ${\bf A}$ does not
necessarily lead to an increase of ${\bf B}$; at some moments of time
${\bf B}$ may actually be reduced by the procedure (8). Moreover, this
procedure can break flux freezing of plasma in the injection region,
by changing the topology of the field there.  However, test
simulations show that none of these effects is serious and the
injected field does have the desired topology.

Magnetic field with poloidal topology is generated in a similar way as
described above, with the same functional form given in equation (8),
except that now only the component $A_\phi$ is taken to be non-zero,
and $A_r$ and $A_z$ are both set to zero.  In this case, the injected
magnetic field lines take the form of poloidal loops around the torus.

In both the toroidal and poloidal cases, the presence of magnetic
field in the injected material allows the redistribution of angular
momentum via magnetic stresses.  As a result, some of the injected
matter loses its initial angular momentum and accretes onto the black
hole, while the excess angular momentum is removed from the
computational domain by the rest of the matter flowing out.  With
time, the initial inner `empty space' becomes filled with accreting
gas, and after several rotation periods as measured at the injection
radius, a quasi-stationary accretion flow is established.  We are
mainly interested in the inner regions of the accretion flow, where
the influence of the outer boundary and the injection region is not
significant. Test simulations show that the dynamics and structure of
the flow inside $\sim R_{out}/2$ depend only weakly on the details of
the injection region.

Absorbing boundary conditions are applied at both the inner and outer
boundaries. The inner boundary consists of a sphere of radius
$R_{in}=2\;R_g$ located inside the marginally stable orbit of the
black hole, while the outer boundary corresponds to the outermost cube
of the Cartesian grid at $R_{out}$.  At both the inner and outer
boundaries, matter is allowed to flow freely out of the computational
domain, but no matter is allowed to return from outside.  Magnetic
field at both boundaries satisfies the divergence-free condition.  The
field at the inner boundary is defined assuming that ${\bf A}={\bf
A}_{init}={\rm constant}$ in the ghost cells located inside $R_{in}$;
here ${\bf A}_{init}$ is the initial (uniform) vector potential.  This
boundary condition on ${\bf A}$ introduces a jump in the magnetic
field near $R_{in}$. However, the jump has no effect on the inflowing
matter because of the strong gravity and super-Alfvenic infall speed.
At the outer boundary, the transverse components of the field with
respect to the sides of the cube are assumed to be zero.  This again
introduces a jump in the field, which results in an artificial torque
on the flow.  The strength of this spurious torque increases with
increasing field strength.

\section{NUMERICAL RESULTS}

The simulations were done on a Cartesian numerical grid consisting of
a number of nested subgrids of increasing resolution inward.  The grid
is designed to provide extra resolution closer to the black hole.  The
number of subgrids varies from 1 to 5 in different runs.  Each subgrid
consists of $64\times 64\times 64$ cells.  The cell size in the
innermost subgrid is $\Delta_1=0.5 R_g$, and each succeeding subgrid
has its cell size increased by a factor of 2.  Therefore, in the
simulations with 5 subgrids, the outermost subgrid has $\Delta_5=8
R_g$, and covers a cube of size $2R_{out}\times 2R_{out}\times
2R_{out}$, where $R_{out}=256 R_g$.  In practice, we use only a
quarter of the full cubic domain, by focusing on a $90^\circ$ wedge
around the $z$-axis. Thus, we employ $32\times 32\times 64$ cells
along $xyz$ and use periodic boundary conditions in the azimuthal
direction.  The nested Cartesian grid has an important technical
advantage with respect to a conventional spherical grid: there are no
polar singularities, and therefore there is no reduction in the
azimuthal grid-size near the pole.  As a result, the time step defined
by the Courant condition is significantly longer compared to a
spherical grid, and for a given amount of computational effort we are
able to run a simulation for a much longer time. On the other
hand, angular momentum conservation is very good in spherical and
cylindrical grids since the azimuthal equation of motion can be
represented in an explicitly conservative form.  A Cartesian grid does
not have this advantage, so angular momentum is likely to diffuse
more.

Spatial re-scaling $r\rightarrow r/R_g$ and time re-scaling
$t\rightarrow t/(R_g/c)$ make the problem independent of the black
hole mass $M$. Further, since we consider non-radiative flows, the
density $\rho$, the pressure terms $P_g$, $Q$, and the electromagnetic
terms $B^2$, $J^2$ in equations (1)-(4) are all proportional to the
mass injection rate, $\dot M_{\rm inj}$.  Therefore, we may eliminate
$\dot M_{\rm inj}$ from the equations and consider only the scaled
solutions. 

When we run a time-dependent simulation using the above code, the
system generally goes through an initial transient phase before
relaxing to a steady state.  Since we are mostly interested in the
final steady state flow, we continue the calculations for a fairly
long time --- long enough that we feel from looking at the results
that the model has reached a steady state.  Even in steady state, the
flow patterns are often time-dependent, so we need to consider the
time-averaged properties of the flow.  Typically, it takes about ten
Keplerian rotational periods, as measured at $R_{out}$, to reach a
steady-state, and we need to average over a few Keplerian periods in
the final state to calculate its average properties.

All our models have been calculated assuming the resistivity parameter
$\eta_0=0.3$. It has been found in previous MHD simulations
(Igumenshchev \& Narayan 2002) that this value of $\eta_0$ is
reasonable.  On the one hand, $\eta_0$ is small enough that it does
not excessively smooth out small-scale magnetic structures.  At the
same time it is large enough to prevent serious loss of energy through
numerical field reconnection.

\subsection{Simulations with Toroidal Injected Field}

The first set of simulations we describe involve the injection of
matter with a toroidal field topology.  These simulations were done
with $A_r=A_\phi=0$ and with $A_z$ equal to the value given in
equation (8).  As explained in \S2.2, this leads to a purely toroidal
injected field.  As the accreting matter moves in, however, poloidal
field of either sign is generated locally, though the total magnetic
flux through the equatorial plane is still zero to within the accuracy
of truncation errors.

We have calculated a range of models with different values of the
parameter $\beta_0$ (which describes the strength of the injected
magnetic field) and with different numbers of nested Cartesian
subgrids.  We find that the flow pattern in steady-state is
practically independent of $\beta_0$.  The magnetic field in the inner
regions of the flow reaches a saturation level which is almost the
same for $\beta_0$ anywhere in the range $1$ to $10^4$.  The reason is
that the magnetic field adjusts near the injection region itself to a
certain saturation level, and maintains that level at all smaller
radii.  Machida et al. (2001) also found that the flow structure is
independent of the initial strength of the toroidal magnetic field.

We discuss here our most advanced simulation, Model~A, which was
calculated on 5 subgrids with an outer radius $R_{out}=256 R_g$ and
injected field parameter $\beta_0=10^2$.  By the end of the
simulation, after about ten orbits at $R_{out}$, Model~A
achieved steady state in a time-averaged sense.  Figure~1 shows a
snapshot of the density distribution in this model in meridional
cross-section.  We see that the accreting matter forms a thick disk,
in which the density is concentrated at the equatorial plane and
increases toward the black hole.  The small-scale fluctuations of
density which are visible are due to turbulent motions.  The amplitude
of the fluctuations is not large.  Figure~2 shows a snapshot of the
pressure distribution, which is seen to have even smaller amplitude
fluctuations compared to the density.  This is because the turbulence
is subsonic.

\placefigure{fig1}
\placefigure{fig2}

The turbulent motions in the gas in Model~A are generated by magnetic
interactions, which provide the main dissipation mechanism.  Most of
the dissipation of energy arises from the resistive term, $\eta{\bf
J}^2/4\pi$, in equation (3); the contribution from the artificial
viscosity is very low since there are no shocks.  During accretion,
the binding energy of matter is efficiently converted into magnetic
energy by two distinct mechanisms: field dragging by the radially
converging flow (Shvartsman 1971), and field amplification by a
magnetic dynamo associated with the magneto-convection and turbulence.
In Fig.~3, we show a schematical example of how the local magnetic
field may be amplified.  A perturbed magnetic line is stretched in the
azimuthal direction by the Keplerian differential rotation, following
which the line self-reconnects, leaving behind a closed magnetic loop
in addition to the original line.  This mechanism represents a
non-linear stage of the magneto-rotational instability (Balbus \&
Hawley 1991).  Field amplification is accompanied by field dissipation
via magnetic reconnection, which releases energy locally at the
reconnection site.  The combined effect of a large number of
reconnection events, randomly distributed in time and over the disk
volume, drives the turbulent motions and fluctuations in the flow.

\placefigure{fig3}

One finds thus a feedback link between turbulence and reconnection:
the turbulent motion tangles magnetic lines and amplifies the magnetic
field, the tangled lines and turbulent motion increase the number of
reconnection events, and the reconnections feed the turbulent motions
through local dissipation of the field.  Obviously, there is a
saturation state in which the rate of amplification of magnetic energy
is balanced by the rate of dissipation via reconnection.

In Fig.~4 we show a typical example of the topology of the magnetic
lines as projected on the equatorial plane.  Here and below, we
use the line integral convolution method (LIC, Cabral \& Leedom 1993)
to illustrate vector fields.  We see that the lines are not simply
stretched tangentially as we would expect for a pure Keplerian flow.
Rather, the lines are strongly perturbed by the presence of many
magnetic loops.  In the meridional section shown in Fig.~5, we again
see that reconnection and turbulent motion of matter result in a very
complicated topology of poloidal magnetic lines.  Especially in the
equatorial regions, we see many magnetic vortices and loops of
different spatial scales.  In contrast, in the narrow polar regions,
the magnetic topology is dominated by a regular field, in which
magnetic lines are oppositely directed in the upper and lower funnels.

\placefigure{fig4}
\placefigure{fig5}

Figure~6 shows a snapshot of the plasma $\beta$ in meridional
cross-section.  We see that $\beta$ is highly inhomogeneous as a
result of the stochastic reconnection and local dissipation of
magnetic field described above. In individual fluctuations, the
magnetic pressure can be up to $\sim 1-2$ orders of magnitude larger
than the average value.  However, the gas pressure $P_g$ fluctuates
very little and $P_m$ never exceeds $P_g$. On average, $\beta\sim
10^2-10^3$ in the equatorial region of the accretion flow, and
decreases towards the poles.

\placefigure{fig6}

The dissipation of magnetic field via reconnection results in an
increase in the entropy of the local gas. Figure~7 shows the distribution of 
the specific entropy in meridional cross-section.  We see that the entropy
increases with decreasing radius in the equatorial plane, and increases
from the equator to the poles at a given radius.  The entropy
distribution shows significant inhomogeneities in the form of tangled filaments.
These can be explained in terms of convective motions in the unstably
stratified medium.  Indeed, the convective motions seen in this MHD
model suggest that the flow is closely related to the CDAF discovered in
viscous hydrodynamical simulations (\S1).  
%As in the case of a CDAF, Model~A shows that 
In particular, Figure~8 shows that in Model~A
the time-averaged surfaces of constant specific entropy are closely
aligned with those of specific angular momentum. This effect, noticed
already for convective thick accretion disks by Bardeen (1970) and
Paczy{\'n}ski \& Abramowicz (1982), 
and found in viscous hydrodynamic simulations by Igumenshchev et al.
(1996) and Stone et al. (1999), indicates that the medium is
marginally stable with respect to the H{\o}iland criterion (Tassoul 1978). The
polar regions of the flow, however, are filled with matter that has a very low
density and high entropy and there is no convection there.  We discuss
the connection between Model~A and the CDAF model in more detail in \S4.1.

\placefigure{fig7}
\placefigure{fig8}

Figure~9 shows a snapshot of the flow streamlines in Model~A projected
on a meridional cross-section. Most of the flow is dominated by
time-dependent vortices of different spatial scales.  Matter accretes
by gradually moving toward the black hole, passing through these
vortices.  Only in the narrow polar directions do the streamlines show
regular inflow into the black hole. However, the accretion rate there
is relatively small because of the low density.  Figure~10 shows the
time-averaged flow pattern, averaged over about three Keplerian
rotation periods measured at $R_{out}$.  The pattern consists of
stationary large scale ($\sim R$) meridional circulation cells.  Thus,
even the time-averaged flow shows deviations from a simple radial
pattern.

\placefigure{fig9}
\placefigure{fig10}

We should emphasize that neither the snapshot in Fig.~9 nor the
time-averaged flow pattern in Fig.~10 exhibits any tendency to produce
a significant bipolar outflow of mass.  In Fig.~10, there is an
equatorial outflow inside a radius $R\simeq 10 R_g$.  Other than this,
there is no evidence for an outflow or a jet, in agreement with a general
theoretical argument by Abramowicz, Lasota \& Igumenshchev (2000).

Figure~11 shows time-averaged radial profiles of selected gas
properties on the equatorial plane.  The upper left panel shows the
density $\rho$.  Leaving aside the spiky outer region near the
injection radius and the innermost region near the black hole, we see
that the density follows a roughly power-law behavior with radius, say
over the range $R \simeq 10-100 R_g$.  The upper right panel shows the
specific angular momentum $\ell$.  We see that the rotation is
super-Keplerian over a small range of radius near the outer boundary,
is sub-Keplerian over much of the intermediate region, and is nearly
Keplerian near the black hole.  The overall behaviour is similar to
that found in global models of  thick disks, slim disks and ADAFs
(e.g., Jaroszy{\'n}ski et al. 1980;
Abramowicz et al. 1988; Narayan, Kato \& Honma
1997; Chen, Abramowicz \& Lasota 1997). The super-Keplerian nature of
$\ell$ near the injection point was discussed by Abramowicz,
Igumenshchev, \& Lasota (1998).  The lower left panel shows the radial
velocity $v_R$, which is found to have a very non-regular
behaviour. This is explained by the meridional circulation patterns
seen in the time-averaged flow shown in Fig.~10.  The spikes in the
plot are due to changes in the sign of the velocity.  The lower right
panel shows the ratio $c_s/v_K$, which characterizes the relative
thickness of the flow: $H/R\simeq c_s/v_K$.  The ratio does not change
much with radius.

\placefigure{fig11}

Over the interior of the flow, between about $10R_g$ and $100R_g$, we
find approximately that $\rho\propto R^{-1}$, $c_s\propto R^{-0.5}$,
and $\ell \propto \ell_K \sim R^{1/2}$.  The radial velocity $v_R$
does not show a very clear power-law behavior.

Figure~12 shows the average total stress $T_{R\phi}$ in meridional
cross-section:
\begin{equation}
T_{R\phi}=\overline{\rho\delta v_R \delta v_\phi}
-\overline{\rho\delta v_{AR} \delta v_{A\phi}} \equiv
T_{R\phi}^R+T_{R\phi}^{mag},
\end{equation}
where $T_{R\phi}^R$ is the Reynolds stress associated with fluid
motions and $T_{R\phi}^{mag}$ is the Maxwell stress associated with
magnetic field fluctuations.  In the definition of the latter,
$\delta{\bf v}_A\equiv\delta{\bf B}/\sqrt{4\pi\rho}$ represents the
fluctuations in the Alfven speed due to field fluctuations.  We find
that $T_{R\phi}$ is nearly axisymmetric.  Figure~12 shows that the
sign of $T_{R\phi}$ behaves non-monotonically over much of the flow.
Regions of negative $T_{R\phi}$ form regular structures which are
extended in the vertical direction and are strongly correlated with
the locations of circulation patterns in Fig.~10.  In particular, there
is an extended region of negative $T_{R\phi}$ at intermediate angles
between the poles and the equator, where angular momentum is
transported inward.  Near the equator, however, the angular momentum
is on average transported outward.  We discuss the behavior of the
stress in greater detail in \S4.1.

\placefigure{fig12}

Simulations on nested Cartesian grids can suffer from artificial
perturbations in the flow at subgrid interfaces. These perturbations
are due to the sudden change in the resolution across subgrids, which
can introduce errors in interpolation. To check the influence of these
errors in our simulations, we have calculated several test models on a
spherical grid. These simulations showed good qualitative and
quantitative agreement with models calculated on the nested Cartesian
grid, indicating that any perturbations due to the presence of
subgrids are weak.  Although the spherical grid has important
advantages on account of its simple and monotonic structure, we found
that it is not practical to use it for the full range of simulations.
This is because models on a spherical grid require significantly
smaller time steps on account of the Courant condition near the poles
(see the discussion at the beginning of \S3) .

\subsection{Simulations with Poloidal Injected Field}

The previous section described simulations with a toroidal injected
field.  Here we describe analogous models with a poloidal field.  In
these simulations, we assumed that the injected gas has a vector
potential with $A_r=A_z=0$ and a non-zero $A_\phi$.  The range of
$\beta_0$ that we have used varies from $10^2$ to $10^4$, which means
that the injected field is always much below equipartition strength.
We find that decreasing $\beta_0$ speeds up evolution of the model to
the steady-state; however, as in the toroidal case, the final
steady-state does not depend on $\beta_0$.  We have simulated several
low resolution models on a nested Cartesian grid, using $1-2$
subgrids.  We refer to our representative model, calculated on 2
subgrids with $R_{out}=64\,R_g$, as Model~B.  Because of the low
resolution of the models, in the following we focus only on
qualitative aspects of the evolution of the models.

The Keplerian differential rotation of the injected material stretches
the poloidal magnetic lines in the azimuthal direction. This leads to
a redistribution of angular momentum and spreading of the matter in
the radial direction. A good fraction of the injected gas moves
outward and leaves the computational domain.  The rest moves inward
and forms an accretion flow.  At the beginning of the simulations, the
magnetic field in the injection torus has zero net $B_z$.  At later
times, however, as the outflowing matter leaves the computational
domain and carries away magnetic field of predominantly one sign of
$B_z$, the inflowing matter drags magnetic field with the opposite
sign of $B_z$ inward.  As a result, the accretion flow acquires a
non-zero net $B_z$.

We find that the evolution of the accretion flow goes through two
distinct stages.  First, there is a transient stage during which the
gas in the injection torus spreads out and forms an accretion disk
stretching down to the black hole.  This transient stage is then
followed by a quasi-steady state in which the magnetic field exerts a
significant back-pressure on the accreting gas.

In the transient stage, the geometry of the flow is very similar to
that described by Hawley, Balbus \& Stone (2001) in their 3D
simulations of non-radiative MHD flows.  The accreting gas forms
a nearly Keplerian disk with a vertically extended low-density corona.
The magnetic field has a simple dipole topology near the outer
boundary, but tends to be complicated at smaller radii, with toroidal
fields of opposite directions above and below the disk.  The field
experiences reconnection events which cause the matter in the
equatorial region to be locally heated up.  Due to the combined
effects of reconnection heat, toroidal magnetic field pressure and the
centrifugal mechanism (Blandford \& Payne 1982), a fraction of the
disk matter is ejected in the vertical directions to feed the low
density corona.  The corona is mostly in dynamical equilibrium, but a
small fraction of the matter in it forms a hot, low density, high
velocity, bipolar outflow.  The outflow carries a significant fraction
of the liberated energy in the form of kinetic energy, heat and
Poynting flux.  As a result, the accretion disk, though quite hot, is
not as hot or thick as in Model~A.

With increasing time, the above transient phase is replaced by a
steady state configuration with very different properties.  The change
occurs because the dipole magnetic field that is dragged inward by the
accretion flow, accumulates in the vicinity of the black hole.  Note
that, while gas can fall into the black hole and disappear, open field
lines cannot vanish.  With time, the accumulated dipole field becomes
quite strong and its influence begins to extend to large radii.  When
the magnetic energy density reaches equipartition with the thermal
energy density of the disk, the magnetic pressure is strong enough to
suppress accretion.  After this time, matter can move toward the black
hole only by `diffusing' through the field, which it does by means of
local interchanges followed by reconnection.  The simulations show
that the accreting matter becomes significantly sub-Keplerian and is
no longer disk-like.  Instead, the matter forms narrow and very slowly
rotating streams, which radially penetrate the highly magnetized low
density medium near the black hole.

With time, the region over which the accumulated field dominates
increases in size until it finally fills the entire computational
domain.  At this point, the model has reached a quasi-steady-state in
which the net magnetic flux no longer changes significantly with time.
Instead, any new magnetic field line which is injected with new matter
is pushed out through the outer boundary by the magnetic pressure.
Matter, however, can decouple from the field, `diffuse' through the
existing magnetic lines along the radial narrow streams mentioned
above and fall into the black hole.

Figure~13 shows the distribution of density in Model~B in this
steady-state, after more than 10 orbits at $R_{out}$.
Four streams are seen as an increase of density in the equatorial
plane.  The accretion velocity in the streams is significantly
sub-sonic and sub-Alfvenic. The fact that there are four symmetric
streams is because we have simulated only a $90^o$ wedge.  A full
$360^o$ simulation is likely to find a different number of streams.
Nevertheless, we believe that the basic result of this calculation,
namely that accretion becomes highly non-axisymmetric, is robust.  The
topology of the magnetic lines in meridional cross-section is shown in
Fig.~14.  We see that the magnetic field is dominated by a strong
dipolar component.  The magnetic $\beta\sim 10^{-3}-10^{-4}$ in the
regions away from the streams, and $\sim 1$ within the streams.
Surfaces of constant total pressure ($P_g+P_m$) are nearly spherical.
Figure~15 shows the structure of the magnetic field projected on the
equatorial plane. The structure is complicated and time-dependent,
reflecting the irregular character of field dissipation in the dense
accreting streams and low density coronal medium.

\placefigure{fig13}
\placefigure{fig14}
\placefigure{fig15}

\section{DISCUSSION}

The results described in \S3 show that the nature of a radiatively
inefficient MHD accretion flow is very different depending on whether
the injected material is dominated by toroidal or poloidal fields.  In
\S\S4.1 and 4.2 below we compare the results we have obtained for
these two limiting cases to our previous work on viscous hydrodynamic
and MHD flows.  Then, in \S4.3, we compare our results to simulations
done by other groups, and in \S4.4, we discuss applicability of the 
results to real accretion flows.

\subsection{Toroidal Field Injection}

Model~A, which involves the injection of matter with a toroidal field
configuration, has a number of distinct features (\S3.1).  Field
amplification and reconnection leads to a significant level of heating
in many localized regions spread all over the flow.  The heating
causes the average entropy to increase as a function of decreasing
radius, as well as to increase from the equator to the pole (Fig.~7).
Since the gradient of the entropy tends to be unstable (by the
H{\o}iland criterion), convective motions are set up which lead to
considerable turbulence in the medium.  We see evidence for this
turbulence in all the diagnostics we have considered: density
(Fig.~1), pressure (Fig.~2), magnetic field (Fig.~4, 5), plasma
$\beta$ (Fig.~6), velocity streamlines (Fig.~9, 10).  The
well-developed turbulent eddies ensure that fluid elements cannot
accrete smoothly down to the black hole from large radius, but rather
must random walk through a series of eddies.  This has the effect of
suppressing the mass accretion rate onto the black hole.

Convection in radiatively inefficient accretion flows, or ADAFs, has
been discussed for several years, mostly in the context of viscous
hydrodynamics.  The possibility of convection in ADAFs was emphasized
by Narayan \& Yi (1994, 1995ab), but the surprisingly strong effect that
convection has on the flow was not appreciated until numerical
simulations were carried out by Igumenshchev \& Abramowicz (1999) and
Stone et al. (1999).  Igumenshchev \& Abramowicz (2000) found that
convection is most important when the viscosity is weak, specifically
when the viscosity parameter $\alpha$ is less than about $0.1$.  This
and other studies also showed that convection has the effect of
suppressing the mass accretion rate onto the black hole.

Using the hydrodynamic simulations as a guide, Narayan et al. (2000)
and Quataert \& Gruzinov (2000) developed an
analytical model describing a new form of radiatively inefficient
accretion called CDAF. The model has two important features.

First, it postulates that convection transports angular momentum
inward rather than outward; that is, the Reynolds stress,
$T_{R\phi}^R$, associated with the fluid fluctuations is negative.
The viscous stress is, however, always outward, i.e.,
$T_{R\phi}^{visc}>0$.  Thus, there is a competition between convection
and viscosity.  In fact, the two fluxes almost cancel each other in a
CDAF so that the net angular momentum flux is much smaller than either
of the fluxes individually: $0 < T_{R\phi}^R+T_{R\phi}^{visc} \ll
|T_{R\phi}^R|, |T_{R\phi}^{visc}|$.

Second, convection drives an outward energy flux $F_c$, which
dominates over other forms of energy transport.  As a result, the flow
is driven to a state of constant convective luminosity with radius,
i.e., $4\pi R^2F_{c}={\rm constant}$.  Narayan et al. (2000) and Quataert
\& Gruzinov (2000) showed that this immediately implies that the
density should scale as $\rho \sim R^{-1/2}$.

Both effects cause the mass accretion rate to be reduced.  Note that
the first of the two effects is specific to rotating flows, whereas
the second can occur in either rotating or non-rotating flows.
Indeed, Igumenshchev \& Narayan (2002) showed that spherical accretion
of a non-rotating magnetized plasma leads to strong convection, which
causes many of the effects seen in rotating CDAFs: constant convective
luminosity, density varying as $R^{-1/2}$, and suppressed mass
accretion.

The presence of well-developed convection in Model~A suggests that
this rotating MHD model may be similar to a CDAF.  For instance,
Model~A has a relative disk thickness, $H/R\simeq 0.5$, a flattened
density profile, and nearly Keplerian rotation, all of which make it
similar to the CDAF model shown for instance in Fig.~17 of
Igumenshchev \& Abramowicz (2000).  There is, however, 
a controversy in the literature on whether or not an MHD flow can behave
like a CDAF.  Balbus \& Hawley (2002), based on linear analysis,
claimed that it is highly unlikely that there could be modes in an MHD
accretion flow that transport angular momentum inward (see also
Christodoulou, Contopoulos \& Kazanas 2003, who come to a similar
conclusion), whereas Narayan et al. (2002) argued that convective
modes in an MHD medium can transport angular momentum inward provided
the mode wavelength $\lambda$ is long enough; specifically, they
required $\lambda/H \gg \beta^{-1/2}$.

Figure~12 shows the distribution of total stress $T_{R\phi}$ in the
meridional plane of Model~A.  We see that the stress is both positive
and negative in different regions, which makes it hard to identify in
which direction angular momentum flows on average.  Figure~16 shows
separately the spherically averaged Reynolds stress $T_{R\phi}^R$
(filled dots) and Maxwell stress $T_{R\phi}^{mag}$ (open circles) as
functions of the radius.  Each quantity has been calculated over
spherical shells of thickness equal to the inner radius.  Thus, the
average over a shell from $R$ to $2R$ is plotted at the geometric mean
radius $\sqrt{2}R$, the average from $2R$ to $4R$ is plotted at
$2\sqrt{2}R$, etc.  In comparing the results to the idealized CDAF
model, it should be noted that the Maxwell stress in an MHD medium
plays the same role as the viscous stress $T_{R\phi}^{visc}$ in a
viscous hydrodynamic flow.

\placefigure{fig16}

Despite the coarse averaging procedure we have used, which corresponds
to heavy smoothing, the results are unfortunately still noisy.  This
is a consequence of the fact that there are large-scale circulation
patterns in the flow which introduce correspondingly large-scale
effects on the stress (e.g., Fig.~12).  Nevertheless, some effects
appear to be clear.  First, we see that the Maxwell stress is always
positive, i.e., the magnetic stress always moves angular momentum
outward, just as viscosity does in the hydrodynamic case.  Secondly,
and more importantly, we see that over a good fraction of the interior
of the flow (i.e., away from the boundaries), the Reynolds stress
tends to be negative, i.e., the convective eddies move angular
momentum inward.  The sum of the two stresses is always positive,
which means that angular momentum is moved outward on average.  These
features are qualitatively similar to what is seen in a hydrodynamic
CDAF and suggest that MHD flows with toroidal field injection are
analogous to the CDAF.  One should keep in mind, however, that the
numerical results are noisy.  In addition, there is not a near
cancellation of the two stresses, as expected in the idealized CDAF
model.  Rather, the Maxwell stress dominates over the Reynolds stress
(as indeed expected by Balbus \& Hawley 2002).  This may be the result
of the relatively coarse resolution and small radial dynamic range of
the present simulations.  Another possible consequence of the net
outward angular momentum flux is that an MHD CDAF might consist of two
distinct parts: an almost non-rotating inner part and a
quasi-Keplerian rotating outer part (Igumenshchev 2002).

As mentioned above, another feature of the CDAF model is that the
density has a shallow radial dependence, $\rho \sim R^{-1/2}$,
compared to the steeper dependence, $\rho \sim R^{-3/2}$, expected in
the absence of convective angular momentum transport (Narayan \& Yi
1994).  Fig.~11 shows that the density in Model~A is indeed shallower
than $R^{-3/2}$, which is consistent with the expectations of the CDAF
model.  However, the profile is not as shallow as the theoretically
expected $R^{-1/2}$, but rather appears to vary as $\sim R^{-1}$.
We note in this context that not all numerical simulations of
CDAFs give $R^{-1/2}$.  Igumenshchev \& Abramowicz (2000), for
instance, found $\rho \sim R^{-0.5}$ for $\gamma=5/3$, but $\sim
R^{-0.7}$ for $\gamma=4/3$.

One possible explanation for the discrepancy in the radial dependence
of the density is that convection moves energy not only in the radial
direction, as assumed in the simple one-dimensional CDAF model
developed by Narayan et al. (2000) and Quataert \& Gruzinov (2000),
but also in the vertical direction.  Imagine dividing the flow into an
equatorial thick `disk' region where most of the accretion occurs and
a polar `corona' region.  In the disk region, apart from convective
energy transport in the radial direction, there could also be vertical
leakage of energy into the corona.  Such a vertical component of the
convective flux has been seen in viscous hydrodynamical simulations
(e.g. Abramowicz, Bj{\"o}rnsson \& Igumenshchev 2000).  Obviously, if
there is a loss of energy from the disk, then the radial convective
luminosity $R^2F_{c}$ within the disk will not be independent of
radius, as assumed in the CDAF model, but will decrease with
increasing radius.  It is easy to see that this will cause the density
profile to steepen relative to the pure CDAF profile.

To be more quantitative, let us consider the same form of the energy
equation as in Narayan et al. (2000), but with an additional cooling
term $Q^{-}$ to describe the vertical loss of energy,
\begin{equation}
\rho v T {ds\over dR}+{1\over R^2}{d\over dR}(R^2 F_c)=Q^{+}-Q^{-},
\end{equation}
where $s$ is the specific entropy, $T$ is the temperature, $F_c$ is
the outward convective energy flux through the disk and $Q^{+}$ is the
energy dissipation rate per unit volume.  In the CDAF solution, one
neglects all the terms except the second one on the left and thereby
finds that the convective luminosity $4\pi R^2F_c$ is independent of
radius.  For the present case, let us retain the cooling term and
write
\begin{equation}
{1\over R^2}{d\over dR}(R^2 F_c)=-Q^{-},
\end{equation}
and let us specify $Q^{-}$ by assuming that a fraction $\xi$ of the
outward flux $F_c$ escapes in the vertical direction:
\begin{equation}
Q^{-}=\xi{F_c\over H}.
\end{equation}
In radiatively inefficient accretion flows the temperature is close to
virial, so the outward convective flux can be estimated to be
\begin{equation}
F_c\propto\rho v_K^3\propto\rho R^{-3/2}.
\end{equation}
Substituting the estimate (13) into equation (11), we find that $\rho$
varies as $R^{-a}$, where
\begin{equation}
a={1\over 2}+{\xi\over H/R}.
\end{equation}
Equation (14) shows that, in the presence of an extra source of
cooling as represented by a value of $\xi>0$, the radial density
profile becomes steeper than the canonical $R^{-1/2}$ of the CDAF
model. Typically, we have $H/R\approx 0.5$.  Thus, in order to explain
the slope of $a\sim1$ found in Model~A, we require $\xi\sim1/4$.
Thus, about 25\% of the local convective energy flux in the disk must
escape into the corona.

It should be noted that, despite the modification described above,
which allows for vertical loss of energy and a range of power-law
indices $a$ for the density, this model is still basically a CDAF,
since it is convection that causes the index $a$ to fall below 3/2.
Another model that has been discussed in the literature (Blandford \&
Begelman 1999) invokes heavy mass loss in an outflow as a result of a
positive Bernoulli constant in the accreting gas (Narayan \& Yi 1994,
1995ab).  This model again leads to a range of values of $a$, but it
is very different in spirit from the CDAF.  In the mass outflow model,
the mass accretion rate in the disk is large at large radii and
decreases as one moves in.  The excess mass is lost from the system in
a powerful mass outflow.  Model~A exhibits significant mass outflow
only in the outermost region, near the radius where mass is initially
injected into the simulation.  Further in, there is practically no
mass lost from the system.  Rather, the gas here participates in
convective eddies.  Any fluid element that temporarily moves outward
as a result of buoyancy turns around after going some distance and
flows back toward the center.  In this sense, Model~A is closer in
spirit to a CDAF than the mass outflow model of Blandford \& Begelman
(1999).

\subsection{Poloidal Field Injection}

Model~B, in which the injected matter has a poloidal magnetic field,
behaves completely differently from Model~A.
In an initial, transient stage, the gas in this model forms a
relatively flattened disk, with a hot corona and a low-density,
high-velocity, bipolar outflow.  There is some qualitative analogy
between this stage of the flow and what has been seen in viscous
hydrodynamic simulations with large viscosity parameter
$\alpha\sim0.3-1$ (Igumenshchev \& Abramowicz 2000).  Those flows too
have powerful unbound outflows.  However, the analogy is not perfect.
For example, the high-$\alpha$ hydro flows have a large accretion
velocity and very sub-Keplerian rotation, whereas the MHD Model~B
described here has a small accretion velocity and nearly Keplerian
rotation.

After the above transient phase, Model~B settles down to a completely
different configuration which is characterized by a strong dipolar
field that fills the entire computational domain.  The field builds up
to this configuration during the transient phase as the accreting gas
advects in magnetic field with the same sign of mean $B_z$.  The
accumulated field imposes strong constraints on the infalling gas.
Once the steady state has been reached, any injected magnetic field
is carried away from the computational domain by a fraction of the
injected gas, while the rest of the gas, minus field, accretes onto
the black hole via long slowly-rotating nearly-radial streams.  

The reason for the buildup of magnetic field can be traced to the
particular prescription we have used for the injected field, in which
the vector potential ${\bf A}$ is assumed to be independent of time.
Because of this, we continuously inject matter with the same sense of
poloidal field.  The gas near the inner radius of the torus always has
one sign of $B_z$ and it is this gas that accretes onto the black
hole, causing a steady buildup of the magnetic flux.  The gas near the
outer radius of the injection torus escapes from the computational
domain, carrying away flux of the opposite sign.

This explains why we find such a profound difference between
simulations with toroidal field injection (Model~A) and poloidal field
injection (Model~B).  In the former case, the injected gas has no
net magnetic flux.  Therefore, even though part of the gas accretes and
part escapes, the part that accretes does not carry a net flux and
there is no opportunity for the flux to build up near the black hole.

Somewhat analogous behavior to that described above for Model B has
been seen in spherical MHD accretion.  Igumenshchev \& Narayan (2002)
simulated Bondi accretion from an external medium in which the
external gas had a uniform magnetic field.  They found that, as the
gas flows in, the field around the black hole builds up, causing a
back-reaction on the accreting gas.  This is one of the reasons
(though the main reason is convection) for causing the accretion rate
to fall.  The main difference between the Bondi simulation and Model~B
described here is the geometry of the injected mass.  In the Bondi
problem, the mass comes in spherically, whereas in Model~B, the mass
is injected in a torus with a disk-like geometry.  With a disk-like
source, there is a large solid angle of empty space through which a
low density outflow can carry away energy and angular momentum.
However, when the incoming mass has spherical symmetry, any gas that
attempts to carry away energy is stopped by incoming gas.  As a
result, it is more efficient for the gas to transport energy via
convective motions and this is what is seen in the Bondi simulation.
As the magnetic field at the center builds up, however, the spherical
symmetry is broken and the accreting matter in the inner regions takes
the form of a doughnut (see Igumenshchev \& Narayan 2002).

\subsection{Comparison with Previous MHD Simulations}

Several groups have recently published results of 2D and 3D MHD simulations
of radiatively inefficient accretion flows.  Machida et al.  (2001)
studied the evolution of a rotating magnetized torus in which the
magnetic field initially is purely toroidal. During the evolution, the
torus forms a turbulent accretion flow with a flattened radial density
profile. The authors state that their flow is similar to a CDAF, and
indeed their simulation has many points of resemblance to our Model~A.
A detailed quantitative comparison of the two models is, however,
difficult since their simulation extends over a limited range of
radius, so that there are large boundary effects.

The study cited above, and several others, have either not used
explicit resistive terms in any of the MHD equations, or used
resistivity only in the induction equation (4) but without accounting
for resistive dissipation in the energy equation (the last term on the
right in eq.[3]).  These simplifications could cause serious
non-conservation of energy: one has a reduction of magnetic energy via
numerical reconnection or resistive dissipation, but the reduction is
not compensated for by a corresponding increase in the thermal energy
(see \S2.1).  Non-conservation of energy in magnetic reconnection will
lead to a suppression of thermal convection (Igumenshchev \& Narayan
2002).

Stone \& Pringle (2001), Hawley et al. (2001) and Hawley \&
Balbus (2002) have presented simulations of MHD nonradiative accretion
flows, starting from an initial torus of matter with a poloidal
magnetic field.  Their simulations are very relevant to our Model~B.
The initial configuration of the field in the simulations is such that
the inner half of the torus (the parts of the torus inside the radius
of maximum pressure) has vertical magnetic flux of one sign, while the
outer half has magnetic flux of the opposite sign.  Initially, the two
halves exactly cancel each other, and so the total vertical magnetic
flux integrated over the computational domain is zero.  However, with
time, the torus spreads in the radial direction, the inner half moving
toward the black hole and the outer half moving outward.  This
spreading causes the regions with opposite signs of magnetic flux to
move apart.  Therefore, as in our Model~B, the inner part of the torus
forms an accretion flow which carries inward a net vertical magnetic flux
of one sign (the same sign as the inner half of the initial torus).
Flux of the opposite sign moves outward and leaves the computational
domain on the outside, again exactly as in Model~B.

The above authors found that the inflowing matter forms a rotationally
supported disk-like accretion flow sandwiched between low-density
coronal regions. The coronal regions form bipolar outflows. The net
magnetic field in the innermost regions of their model has a bipolar
structure with non-zero net vertical flux (see Figs 4 and 5 of
Stone \& Pringle 2001).  These features are similar
to what we see in our Model~B during its initial transient stage.
However, there are two important details in which the models of Hawley \&
Balbus (2002) differ from our simulations.  They report finding
a small hot torus near the marginally stable orbit and a
magnetically-confined jet near the rotation axis.  These are not
present in Model~B.  An even more important difference is that Hawley
et al. (2001) and Hawley \& Balbus (2002) do not see any accumulation
of magnetic field in the vicinity of the black hole.  Consequently,
their simulations do not exhibit the kind of steady state
configuration we have described in \S3.2 in which the field dominates
the dynamics and gas is able to accrete only via narrow radial
streams.

We believe that this discrepancy is the result of using different
inner boundary conditions.  We use a spherical inner boundary
condition centered on the black hole (see \S 2.2).  Thus, the black
hole is the only absorbing entity, and the $z$-axis away from the
black hole is a non-absorbing regular region of the flow.  In
contrast, Hawley and collaborators use a cylindrical inner
boundary condition, in which the region within $1.5 R_g$ cylindrical
radii of the $z$-axis is removed from the computational domain and
absorbing boundary conditions are applied along this entire excised
region.  While the two boundary conditions give similar results for
hydrodynamic simulations, they cause significant differences for MHD
flows.

Figure~17 schematically shows the difference between the two boundary
conditions.  When magnetized gas flows in, the net magnetic flux in
the accretion flow is conserved if the spherical boundary
condition is used.  Thus, matter falls into the black hole, but open
field lines do not disappear.  In contrast, with the cylindrical
boundary condition, the component of the magnetic field parallel to
the axis of the cylinder gets absorbed, and the simulation does not
conserve the net magnetic flux. Whether or not a magnetic field
line disappears entirly from the grid depends on the velocity
streamlines of the accreting gas.  For instance, under certain special
conditions a field line may remain stuck in the configuration 2 shown
in the left panel of Fig. 17.  However, even in this case, the region
in the vicinity of the black hole loses the magnetic pressure
associated with the advected field line and will behave differently
compared to the spherical boundary case.  This is a serious problem,
especially for accretion flows such as Model~B.  In our simulation of
Model~B, the accumulated field remains in the computational volume and
has a profound effect on the time evolution of the accretion flow.
With the cylindrical boundary condition, however, the field
never has a chance to accumulate since it keeps disappearing at the
axis.  Therefore, one obtains an incorrect description (in our view)
of MHD processes in the vicinity of the black hole.

\placefigure{fig17}

The cylindrical inner boundary condition is convenient and
allows one to avoid certain technical problems near the rotation axis
in 3D MHD simulations.  Perhaps for this reason, a number of recent
numerical simulations of MHD accretion flows around black holes have
employed this boundary condition (e.g. Hawley 2000, 2001; Hawley \&
Krolik 2001, 2002; Armitage, Reynolds, \& Chiang 2001; Krolik \&
Hawley 2002).  The effect of the inconsistency pointed out above on
these simulations needs to be investigated.

2D numerical simulations do not usually suffer from this problem, so
one would expect the magnetic flux accumulation that we find in Model
B to be apparent in such work.  The simulations of Stone \& Pringle
(1999) are interesting in this respect.  Although they start with an
initial torus with a fixed amount of matter rather than a torus with
continuously injected mass as in our work, nevertheless, at late times
when much of the mass has accreted, their simulation ought to exhibit
the magnetic field accumulation at the center that we find.
Unfortunately, it appears that their calculations were not run long
enough.  There is some indication in Fig. 6 of their paper that the
magnetic pressure is becoming dynamically important at small radii.
This is similar to the effect that we see in our Model B, except that
our effect is much larger and more dramatic.

Recently, Casse \& Keppens (2002) published results of 2D numerical
simulations of a magnetized accretion disk with a polytropic equation
of state.  The authors assumed continuous injection of matter at the
outer boundary, which makes their work closer to our simulation
compared to the Stone \& Pringle work.  Unfortunately, they have
evolved their model for a very short time, less than an orbital period
at the injection radius.  Their results are consistent with those of
Hawley and collaborators and also with our results on Model~B during
the early transient phase.  However, since they have not run their
model long enough, they are not in a position to confirm the late-time
field-dominated steady state that we find.

After the present paper was submitted to the journal, Proga \&
Begelman (2003) posted a paper describing 2D MHD simulations of
slowly-rotating radiatively inefficient gas on a black hole.  They
introduce a new magnetic field topology for this problem, namely a
split monopole configuration.  While there are interesting points in
their work, we note that the split monopole field is rather artificial
--- by construction, it always has a fixed magnetic flux!  Thus, there
is no possibility for the field near the black hole to build up over
time.

Since the unusual steady-state that we find for Model B is the most
surprising (and possibly controversial) result of the present paper,
we take this opportunity to summarize the situation:

\noindent
1. In the present work, we have run a 3D MHD simulation (Model B) for
a long time, using a poloidal field configuration in the injected gas
such that the portion of the material that accretes always carries in
the same sign of $B_z$.  We find that the magnetic flux accumulates,
initially near the black hole and then farther out, and the field
becomes dynamically dominant.  At late time, mass is able to accrete
only via narrow streams, in a highly non-axisymmetric manner.

\noindent
2. Igumenshchev \& Narayan (2002) carried out 3D MHD simulations of
spherical accretion of gas with a uniform $B_z$.  Their simulation
showed evidence for field accumulation at the center.  If the
simulation were run for a longer time, we predict that the field would
dominate the entire computational domain, and would have effects
similar to what we see in Model B.

\noindent
3. Hawley et al. (2001) and Hawley \& Balbus (2002) have carried out
3D MHD simulations of rotating flows, but do not see any evidence for
field accumulation at the center.  We argue that their boundary
condition (see Fig. 17) does not conserve magnetic flux.  Therefore,
in our view, their simulations will not exhibit field accumulation
near the black hole even if run for a very long time.

\noindent
4. Stone \& Pringle (1999) have run 2D MHD simulations, starting from
an initial magnetized torus.  There is a slight hint that their final
state has some field accumulation at the center.  If their simulation
were run for a longer time, such that most of the gas either accretes
or is ejected from the computational domain, then we believe that the
final configuration would correspond to the field-dominant state found
in Model~B.  There would be one difference, however.  Since the Stone
\& Pringle simulation is in 2D, it cannot show the kind of
non-axisymmetric accretion via streams that we see in Model B.

\noindent
5. Casse \& Keppens (2002) have carried out 2D simulations with
continuous mass injection in a torus.  If these simulations are run
for a long time, they should resemble Model B.  However, being 2D,
they again will not find the accretion streams that we see in Model B.

\noindent
6. Finally, Proga \& Begelman (2003) have done 2D simulations with a
very special magnetic field geometry (split-monopole) that forces the
magnetic flux to remain constant in time.  This simulation will not
reproduce the flux accumulation that we see in Model B regardless of
how long it is run.

\subsection{Is Model B Relevant for Real Accretion Flows?}

Leaving aside the conflicting results from different numerical
simulations, which we have attempted to sort out in the previous
subsection, the physical concepts behind the long-term evolution of
Model B appear to be sound.  Namely, if gas accretes on a black hole
with magnetic field of a single sign of $B_z$, then magnetic flux will
accumulate around the hole and will, over time, become dynamically
important.  The field will then disrupt the accretion flow and will
cause the gas to flow in via non-axisymmetric streams or blobs, which
will almost certainly have observational consequences.  But how
relevant are our results for real accretion flows?

The particular geometry of injected matter that we have considered in
Model~B is somewhat artificial.  We have assumed that the injection
occurs in a narrow torus, with a poloidal field configuration that is
so designed that the inflowing and outflowing gas have opposite signs
of $B_z$.  This particular geometry is unlikely to occur in nature.
However, the final steady state that we see in Model B does not depend
on the precise geometry.  All that is required is that the accreting
gas should have a net $B_z$ of a single sign (positive or negative)
over an extended period of time. This is by no means unlikely.

In a real accretion flow we imagine that $B_z$ will fluctuate with
time.  If the fluctuations are slow so that the sign of $B_z$ remains
the same for much longer than the accretion time, then the flow ought
to resemble what we see in Model~B.  There should be a transient
period during which the gas would form a disk and would have an
energetic outflow, followed by a long period of time during which the
field would be dominant over a central volume near the black hole and
accretion would be restricted to narrow streams.  When the sign of the
field finally reverses (assuming it does --- see below), there will
again be a transient disk-like phase with ejection in an outflow,
etc., and the cycle will repeat.  During the transient stages, one
would expect to observe
%powerful 
flare activity due to the
annihilation of oppositely directed poloidal fields.  
Note that in this scenario the flare activity could be very
powerful and easily detectable in observations, because stored energy
of dissipated magnetic field is comparable to the binding energy
of accretion matter.
If the poloidal
field reverses its direction frequently, then we imagine that the
accretion flow will resemble the initial transient phase of Model~B at
all times.  Both cases could be investigated by repeating our
simulations with a time-varying $A_\phi$ in the injected gas.

The long-term evolution of the magnetic field near the black hole
depends on the statistics of the magnetic field in the externally
supplied gas.  If the $\Delta B_z$ that is added with each new piece
of matter is completely uncorrelated with previous injections, then
the accreted magnetic flux would random-walk in time, and would on
average increase as $t^{1/2}$.  Alternatively, if the accreting matter
is supplied by an external medium that experiences MHD turbulence with
a large outer scale, then one expects the flux to grow to a large
value and to remain at that value for a long time (as long as it takes
gas to accrete from a radius of order the outer scale).  If, however,
the turbulent spectrum is dominated by small-scale field, then the
flux would not grow very much and the vertical field component would
always be sub-equipartition.  In this case one might expect the
formation of a CDAF-type flow as in Model~A.  In the case of accretion
in binary stars, the statistics of the magnetic field in the outer
envelope of the mass-donor star will determine how much field
accumulates around the primary.  This is poorly understood.

Note that, in most accretion flows, we expect the magnetic field
strength in the disk to be amplified via the magnetorotational
instability, but the additional field so created will on average have
zero $B_z$.  Therefore, it is only the original $B_z$ supplied with
the gas that is relevant for determining how much flux will accumulate
around the accreting star.

The very unusual effects we have seen in Model B, viz., dynamically
dominant magnetic field and accretion via narrow slow-rotating streams,
may not even be restricted to radiatively inefficient accretion flows.
A similar buildup of field and corresponding modification of gas
dynamics could occur also in radiatively efficient accretion flows.

High-energy observations frequently suggest the simultaneous presence
of hot nearly relativistic matter and `cold' quasi-thermal gas in the
inner regions of active galactic nuclei (AGN) and black hole X-ray
binaries.  Magnetic fields are believed to be a common ingredient in
these systems, and it has been proposed that the field may effectively
confine or otherwise couple to clumpy cold gas (Krolik 1998; Celotti
\& Rees 1999).  Model~B exhibits some of the features postulated in
these proposals.  Accretion occurs mostly as relatively cold gas, but
the binding energy of this gas is released through magnetic
reconnection and a large fraction of the heat goes into the hot
phase. X/$\gamma$-rays radiated by reconnecting magnetic flares in the
hot phase, and reprocessed optical-UV radiation from the cold phase,
may explain some of the spectral signatures observed in AGN and X-ray
binaries.

\section{SUMMARY}
 
We have described in this paper global time-dependent
three-dimensional MHD simulations of radiatively inefficient accretion
flows onto black holes.  The simulations are done within the framework
of a pseudo-Newtonian potential to mimic relativistic effects near the
black hole.  The flows extend from an inner radius $R_{in}=2 R_g$ to
an outer radius $R_{out}$, which in different models ranges from
$16R_g$ to $512R_g$.
Magnetized matter is continuously injected into the
computational domain within a torus whose major radius is slightly
less than $R_{out}$ and whose minor radius is about one grid cell.
The material starts off with Keplerian rotation and no initial radial
velocity.  However, matter spreads under the influence of magnetic
torques, and part of the injected matter accretes onto the black hole
while the rest escapes from the computational box.  The following is a
summary of the main results.

1. We have studied two limiting cases for the geometry of the injected
magnetic field: pure toroidal field and pure poloidal field.  We find
that the results are completely different in the two cases.

2. When the injected field is primarily toroidal, which corresponds to
the simulation designated Model~A (see \S3.1), 
the accreting gas forms a nearly
axisymmetric, quasi-stationary, geometrically thick disk with little
mass outflow in a wind or a jet.

3. The magnetic field achieves a saturation level with a plasma
$\beta\sim10^2$.  In this saturated state, field amplification by
radial convergence and dynamo action is balanced by field dissipation
through reconnection.

4. There is considerable convective activity in the flow, driven by an
unstable entropy gradient that results from energy dissipation via
field reconnection.

5. There is a close analogy between our MHD Model~A and the CDAF model
that was developed to explain viscous hydrodynamic simulations of
radiatively inefficient accretion flows.  In particular, there is
evidence that the Reynolds stress in the interior of the flow (away
from the boundaries) moves angular momentum inward rather than
outward.

6. The radial density profile is relatively flat, $\rho \sim R^{-1}$,
but not as flat as the $\rho \sim R^{-1/2}$ predicted by the CDAF
model.  We suggest that the difference is because of the vertical
leakage of convective energy flux from the thick disk (see \S4.1).

7. These results on the toroidal field case are consistent with some
previously reported work in the literature, but our simulations have a
larger dynamic range and have been run for a longer time.

8. When the injected field is primarily poloidal, which corresponds to
the simulation designated Model~B (see \S3.2), 
the accretion flow goes through two distinct phases.  

9. In an initial transient phase, the accreting gas forms a relatively
flattened axisymmetric disk with a hot corona and a bipolar outflow.
This stage is similar to results found by other groups.

10. With time, as the accreting gas continues to bring in magnetic
flux, the magnetic field builds up around the black hole and farther
out.  The strong field disrupts the axisymmetry of the disk, leading
to a completely different steady state flow configuration.

11. In steady state, there is a two-phase medium.  Most of the volume
is filled with a strong dipolar magnetic field with $\beta\ll1$, from
which a thermal wind flows out along the open magnetic lines.  
Accretion occurs mainly via narrow
slowly-rotating radial streams, with $\beta\sim1$, which `diffuse' through
the field.

\acknowledgments

We thank Eliot Quataert and the referee, Jim Stone, for many helpful
comments on the manuscript.  IVI was supported by the U.S. Department
of Energy (DOE) Office of Inertial Confinement Fusion under
Cooperative Agreement No. DE-FC03-92SF19460, the University of
Rochester, the New York State Energy Research and Development
Authority, and RFBR grant 00-02-16135.  RN was supported by NSF grant
AST-9820686.

%\appendix

\clearpage

%\newpage

\clearpage

\begin{figure}
%\plotone{f1.eps}
\caption{Snapshot of the density distribution in Model~A in the $x-z$
plane (meridional cross-section).  The box shown has a size of
$128\times 128\,R_g$, with the black hole at the center.  Magnetized
gas with a toroidal field is injected continuously within a torus at
radius $\simeq 510R_g$ (well outside the box shown here).  The color bar on
the side indicates the scale for $\log\rho$ (in arbitrary units).  
Note that the gas is
concentrated toward the equatorial plane (the horizontal axis) and
increases toward the black hole.  The fluctuations in the density are
caused by convective motions. The two vertical polar funnels are
filled with low-density matter.
\label{fig1}}
\end{figure}

\begin{figure}
%\plotone{f2.eps}
\caption{Snapshot of the pressure distribution in Model~A in
meridional cross-section (see Fig.~1 for details).  The color bar
gives the scale for $\log P$ (in arbitrary units).  
In the two vertical polar funnels, the
pressure is very small as there is very little matter there.
\label{fig2}}
\end{figure}

\begin{figure}
%\plotone{f3.eps}
\caption{Schematic evolution of a perturbed magnetic field line in a
differentially-rotating accretion flow.  The unperturbed velocity is
assumed to correspond to a Keplerian rotation profile.  (a), (b), (c),
(d) show the geometry of a magnetic line as a function of time. (a)
shows a small initial perturbation of the line; (b), (c) the perturbed
segment of the line is stretched by the Keplerian shear, and a
magnetic loop is formed; (d) the field line reconnects and the
magnetic loop separates from the line, causing a net amplification of
the field strength relative to (a).
\label{fig3}}
\end{figure}

\begin{figure}
%\plotone{f4.eps}
\caption{Snapshot of magnetic lines in Model~A in the $x-y$ plane
(equatorial cross-section).  The box shown has a size of $64\times
64\,R_g$, with the black hole at the center.  The component of field
lines parallel to the plane is shown.  Note the large number of
magnetic loops, stretched in the azimuthal direction by the Keplerian
rotation.  The origin of the loops is schematically explained in
Fig.~3. 
\label{fig4}}
\end{figure}

\begin{figure}
%\plotone{f5.eps}
\caption{Snapshot of magnetic lines in Model~A in meridional
cross-section for a box of size $64\times 64\,R_g$.  The component of
field lines parallel to the plane is shown.  Except for the polar
regions, the magnetic field elsewhere has a highly tangled
morphology. This is the result of convection.
\label{fig5}}
\end{figure}

\begin{figure}
%\plotone{f6.eps}
\caption{Snapshot of the distribution of the plasma $\beta\equiv
P_g/P_m$ in Model~A in meridional cross-section for a box of size
$128\times 128\,R_g$. The color bar gives the scale for $\log\beta$.
A complicated pattern is seen in the distribution of $\beta$.  Regions
of large $\beta$, or weak magnetic field, correspond to regions of
reconnection and dissipation of magnetic energy. In the polar regions,
where magnetic field has a more regular structure (see Fig.~5), the
distribution of $\beta$ is more homogeneous.
\label{fig6}}
\end{figure}

\begin{figure}
%\plotone{f7.eps}
\caption{Snapshot of the distribution of specific entropy $s$ in
Model~A in the meridional plane for a box of size $128\times
128\,R_g$. The color bar gives the scale for $\log s$.  Note the
complicated filamentary structure in the equatorial region and in the
intermediate region between the poles and the equator.  This is the
result of field dissipation and convection.  The two vertical polar
funnels are filled with high-entropy low-density matter.
\label{fig7}}
\end{figure}

\begin{figure}
%\plottwo{f8a.eps}{f8b.eps}
\caption{Time-averaged distribution of specific entropy $s$ (left) and
specific angular momentum $\ell$ (right) in Model~A in meridional
cross-section.  The averaging has been done over about three Keplerian
rotation periods measured at $R_{out}=256 R_g$.  Comparing the two
plots, one sees that lines of constant $s$ are closely aligned with
those of constant $\ell$ over much of the volume, with the exception
of the two funnels.  This indicates that the accreting gas is
marginally stable to convection according to the H{\o}iland criterion.
Some local features are evident in the distributions of $s$ and $\ell$
in the equatorial region.  These are due to the circulation patterns
shown in Fig.~10.
\label{fig8a}}
\end{figure}

\begin{figure}
%\plotone{f9.eps}
\caption{Snapshot of velocity streamlines in Model~A in meridional
cross-section, for a box of size $64\times 64\,R_g$.  The component of
streamlines parallel to the plane is shown.  Except for the polar
regions, the flow pattern consists of a number of vortices and eddies,
the result of convective turbulence.
\label{fig9}}
\end{figure}

\begin{figure}
%\plotone{f10.eps}
\caption{Same as in Fig.~9, but for the time-averaged flow.  The
averaging has been done over about three Keplerian rotation periods
measured at $R_{out}=256 R_g$.  Instead of the large number of eddies seen in
Fig.~9, one now sees large-scale meridional circulation patterns which
are quite symmetric with respect to the equatorial plane.  In the
polar regions and in the vicinity of the black hole the streamlines
are directed inward. In the innermost region, inside $10\,R_g$,
outflows are observed, but only near the equatorial plane.
\label{fig10}}
\end{figure}

\begin{figure}
\plotone{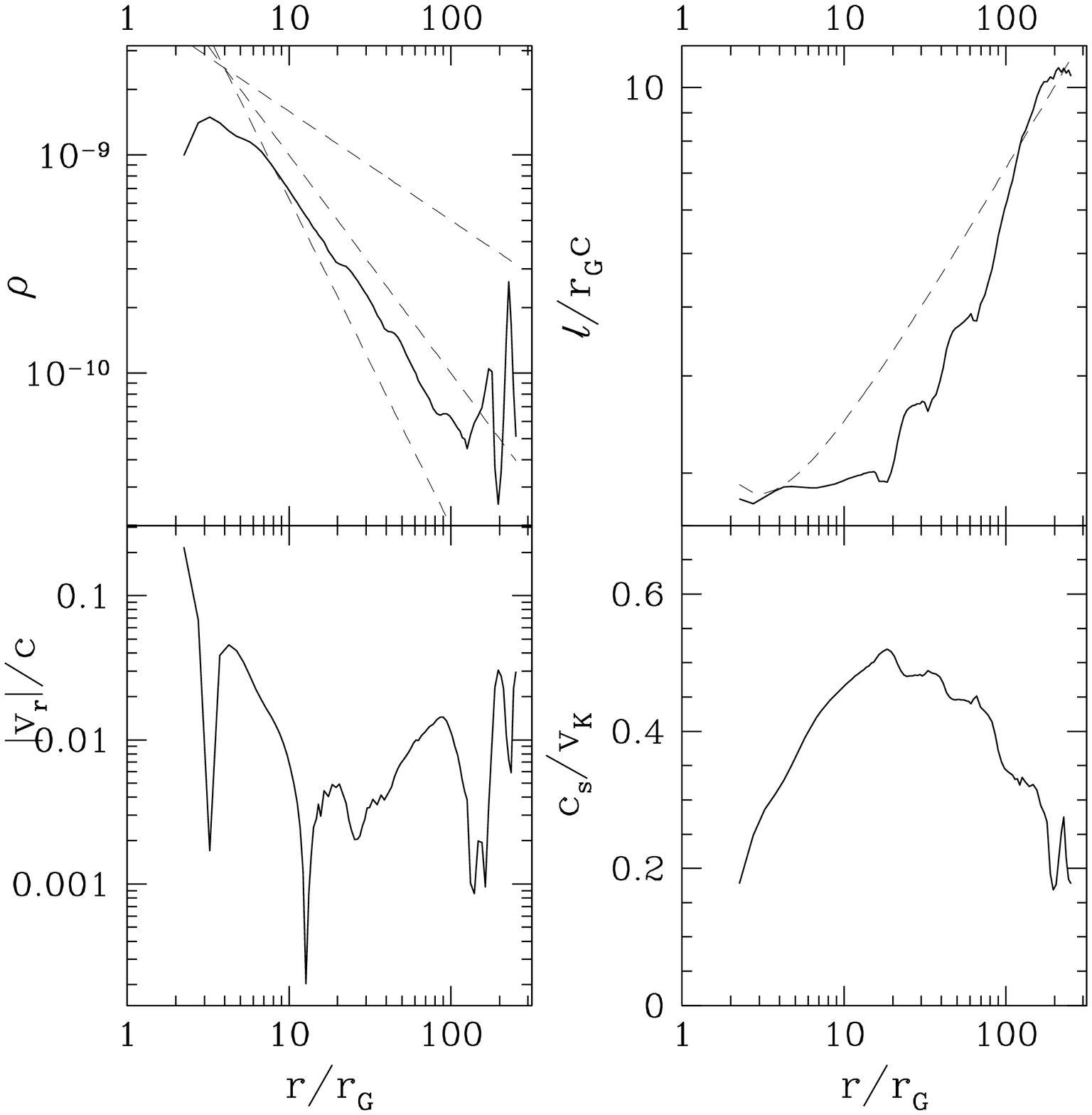}
\caption{Radial structure of the flow in Model~A on the equatorial
plane. All plotted quantities -- density $\rho$ (in arbitrary units), 
specific angular
momentum $\ell$, radial velocity $v_r$, and sound speed velocity $c_s$
-- have been averaged over about three Keplerian periods measured at
$R_{out}=256 R_g$.  The density distribution is shown together with three
dashed lines, which represent power-laws $\propto R^{-a}$ with
$a=1/2$, $1$, $3/2$.  The specific angular momentum is shown along
with a dashed line which represents the Keplerian distribution of
$\ell$ for the pseudo-Newtonian potential (eq.~6) used in the
simulation.
\label{fig11}}
\end{figure}

\begin{figure}
%\plotone{f12.eps}
\caption{Time-averaged distribution of the $R\phi$-component of the
stress tensor, $T_{R\phi}$ (eq.~[9]), in Model~A in meridional
cross-section for a box of size $128\times 128\,R_g$. The color bar
gives the scale for the stress in units of the local time-averaged gas
pressure.  The black line separates regions of positive and negative
stress (positive/negative stress correspond to outward/inward angular
momentum flux, respectively).  Note the correlation between the
locations of negative and positive stresses and the time-averaged
circulation pattern in Fig.~10 (remembering the different spatial
scales in the two figures).
\label{fig12}}
\end{figure}

\begin{figure}
%\plotone{f13.eps}
\caption{Density distribution in Model~B in the $x-y$ plane
(equatorial cross-section) at late times (steady state).  The black
hole is located at the center of a box of size $64\times 64\,R_g$.  At
this stage of the simulation, accretion occurs along almost radially
directed streams moving in toward the black hole. One can see four
symmetric streams in the image, which is the result of the simulation
being done over a quarter of the domain, with azimuthally periodic
boundary conditions (see \S3).
\label{fig13}}
\end{figure}

\begin{figure}
%\plotone{f14.eps}
\caption{Magnetic lines in Model~B in the $x-z$ plane (meridional
cross-section).  The black hole is located at the center of a box of
size $64\times 64\,R_g$.  The component of field lines parallel to the
plane is shown.  The magnetic field clearly has a bipolar structure.
The magnetic field is in rough equipartition with the accretion
streams shown in Fig.~13.
\label{fig14}}
\end{figure}

\begin{figure}
%\plotone{f15.eps}
\caption{Magnetic lines in Model~B in equatorial cross-section for a
box of size $64\times 64\,R_g$.  Note the complicated structure of the
field, which is loosely correlated with the positions of the accretion
streams seen in Fig.~13.
\label{fig15}}
\end{figure}

\begin{figure}
\plotone{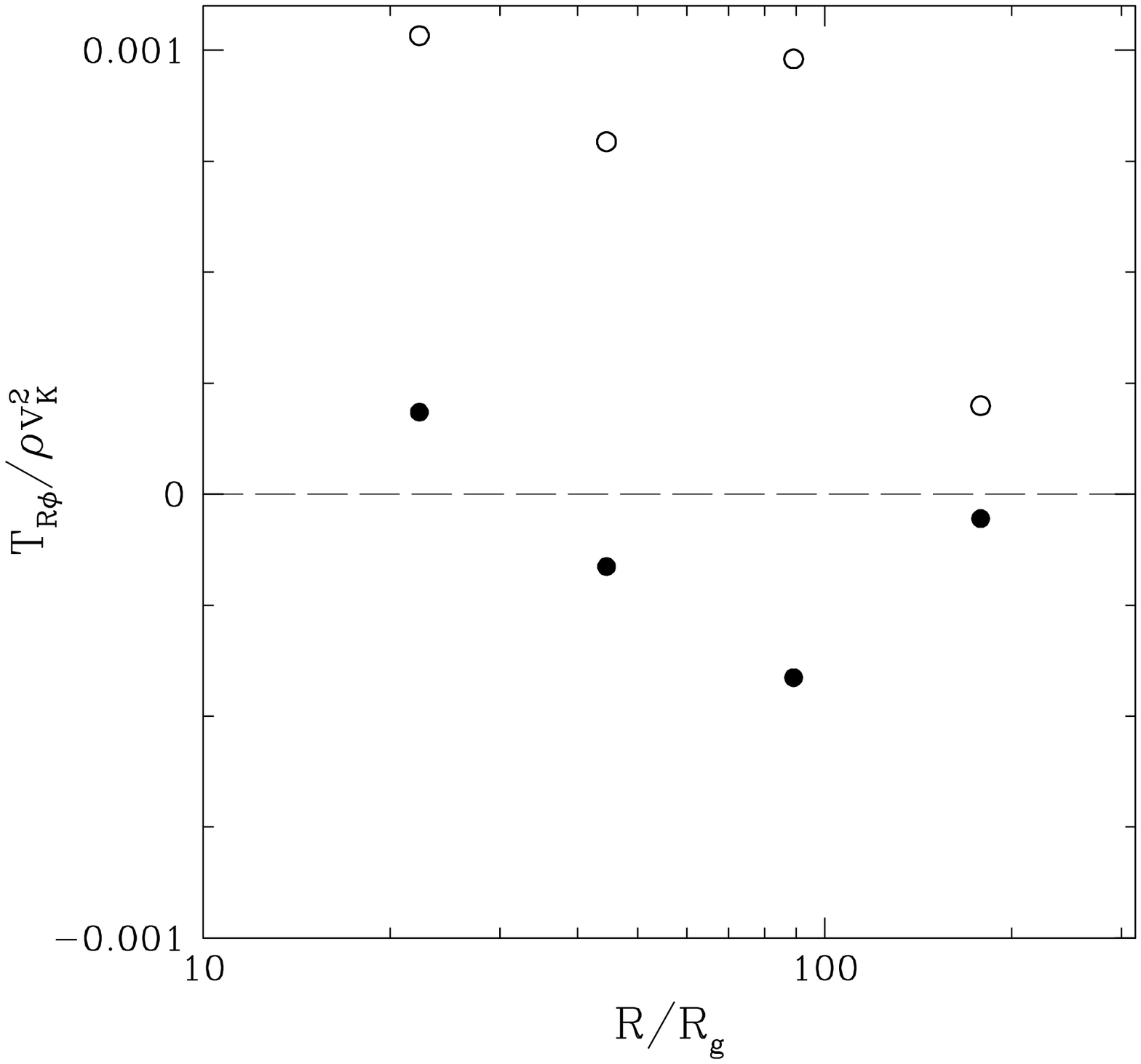}
\caption{Radial distribution of Reynolds stress $T_{R\phi}^R$ (filled
dots) and Maxwell stress $T_{R\phi}^{mag}$ (open circles) in
Model~A. The stresses have been averaged over spherical shells as
described in the text.  Note the negative sign of the Reynolds stress
over a range of radius.  In this region, convection transports angular
momentum inward, in agreement with the fundamental postulate of the
CDAF model. The positive sign of the Maxwell stress indicates that the
magnetic field transports angular momentum outward. The total stress
(sum of Reynolds and Maxwell stresses) is positive, as expected, so
that on average the angular momentum is transported outward.
\label{fig16}}
\end{figure}

\begin{figure}
\plotone{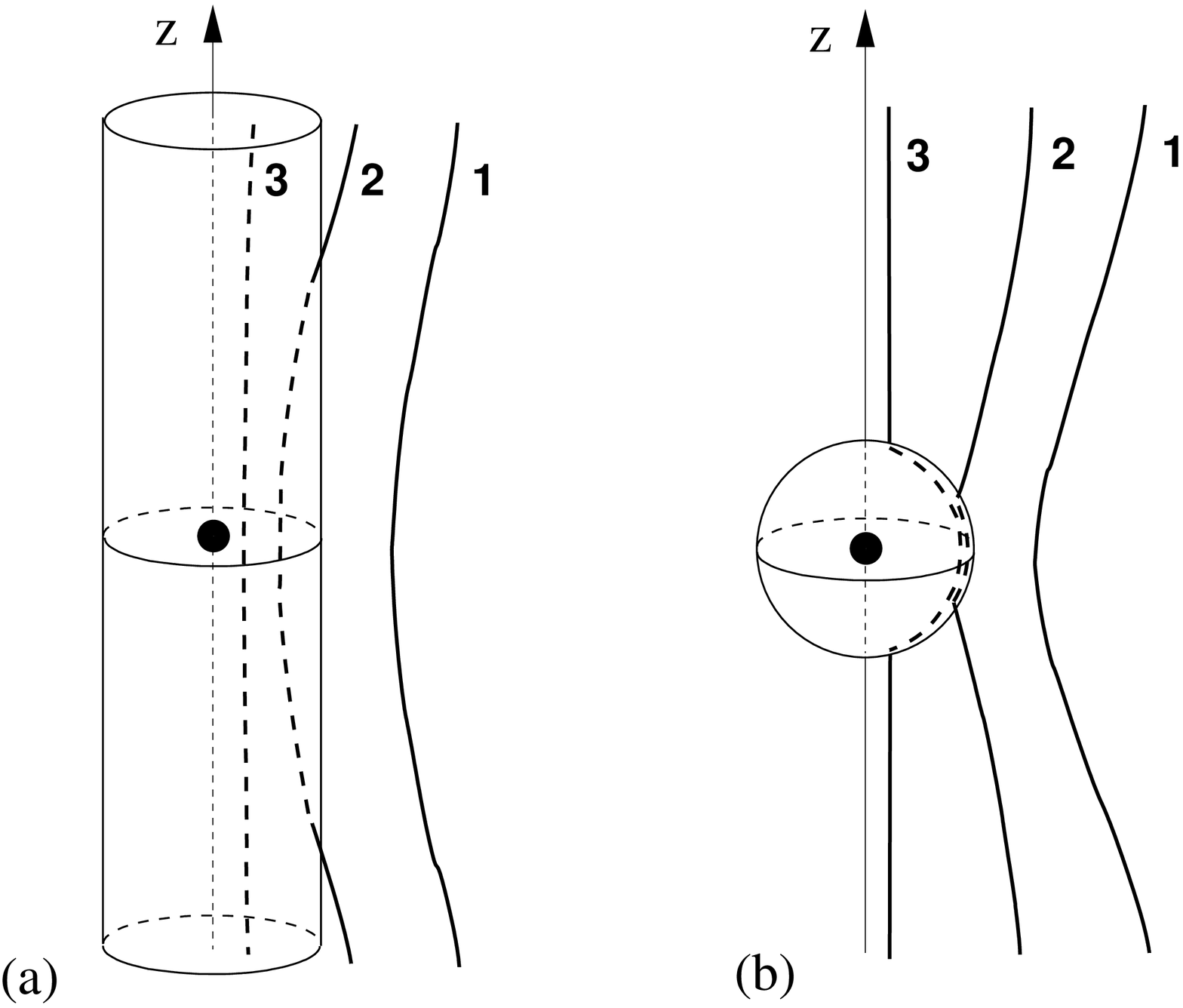}
\caption{Cartoon depiction of the difference between two inner
boundary conditions.  Panel (a): shows the cylindrical boundary
condition used by Hawley et al. (2001) and Hawley \& Balbus
(2002). The cylinder goes through the whole computational domain,
starting at the top outer boundary and ending at the bottom boundary.
Panel (b): shows the spherical boundary condition used in the
present study.  The cylinder in panel (a) and the sphere in panel (b)
indicate the locations of the absorbing boundaries in the two cases.
Any matter or magnetic field that crosses the absorbing boundary stops
contributing to the dynamics of the flow in the computational domain,
thus mimicking the effect of gravitational capture of matter by the
black hole. The bold lines 1, 2, 3 in each panel show the schematic
time evolution of a frozen-in magnetic line as it is advected inward
by the accretion flow. For simplicity, we do not consider the effects
of rotation. In panel (a), the component of the magnetic field
parallel to the $z$-axis is absorbed.  Therefore, with time, a field
line will either disappear entirely from the computational domain
(e.g., line 3 which is entirely inside the absorbing cylinder) or in
rare cases remain partially inside the grid (line 2).  In either case,
the simulation does not conserve the net magnetic flux in the vicinity
of the black hole.  In panel (b), on the other hand, open field lines
do not disappear when the accreting matter falls into the black hole.
Therefore, the net magnetic flux is conserved.  This is shown by lines
2 and 3, which continue to contribute to the magnetic pressure around
the black hole even after parts of the line have moved inside the
absorbing sphere.
\label{fig17}}
\end{figure}

\end{document}